\documentclass[12pt]{iopart}
\usepackage{graphicx,epsfig,colordvi,cite}
\def\be{\begin{eqnarray}}
\def\ee{\end{eqnarray}}

\def\dsp{\displaystyle}
\def\prt{\partial}

\def\lsim{\stackrel{\scriptstyle <}{\phantom{}_{\sim}}}
\def\gsim{\stackrel{\scriptstyle >}{\phantom{}_{\sim}}}
\begin{document}
\title{Catalytic $\phi$ meson production in heavy-ion collisions}
\author{E~E~Kolomeitsev$^1$  and B~Tom\'a\v{s}ik$^{1,2}$}
\address{$^1$Matej Bel University,  SK-97401 Bansk\'a Bystrica, Slovakia }
\address{$^2$Czech Technical University in Prague, FNSPE, CZ-11519 Prague 1, Czech Republic}
\ead{kolomeitsev@fpv.umb.sk}
\begin{abstract}
The phi meson production on hyperons, $\pi Y\to\phi Y$ and anti-kaons $\bar{K}N\to \phi Y$ is
argued to be a new efficient source of phi mesons in a nucleus-nucleus collision. These reactions
are not suppressed according to Okubo-Zweig-Izuka rule in contrast to the processes with
non-strange particles in the entrance channels, $\pi B$ and $BB$ with $B=N,\Delta$. A rough
estimate of the cross sections within a simple hadronic model shows that the cross sections of
$\pi Y\to\phi Y$ and  $\bar{K}N\to \phi Y$ reactions can exceed that of the $\pi N\to \phi N$
reaction by factors $~50$ and $~60$, respectively. In the hadrochemical model for nucleus-nucleus
collisions at SIS and lower AGS energies we calculate the evolution of strange particle
populations and phi meson production rate due to the new processes. It is found that the catalytic
reactions can be operative if the maximal temperature in nucleus-nucleus collisions is larger than
130~MeV and the collision time is larger than 10~fm. A possible influence of the catalytic reactions
on the centrality dependence of the $\phi$ yield at AGS energies and the $\phi$ rapidity
distributions at SPS energies is discussed.
\end{abstract}
\section{Introduction}
The $\phi$ meson production is an important part of the study program of different nucleus-nucleus
collision experiments in the whole range of collision energies: at the
AGS~\cite{AGS-Akiba96,AGS-Back04}, the
SPS~\cite{NA49-Afanasiev00,NA50-Alessandro03,NA60-DeFalco06,CERES-Adamova06,NA49-Alt08},
RHIC~\cite{STAR-Adams05,PHENIX-Adler05}, and even at deeply subthreshold energies at the
SIS~\cite{FOPI-Mangiarotti03}. As the longest-living vector meson, the $\phi$ is considered
to be a good probe of the collision dynamics: it would decay mainly outside the fireball and the
daughter hadrons would be weakly affected by rescattering. The dominant hadronic
decay is $\phi\to \bar{K}\,K$, hence, the production of $\phi$'s interplays with the
production of strange mesons in nucleus-nucleus collisions and can provide a complementary
information on the strangeness dynamics. As other neutral vector mesons, the $\phi$ contributes to
dilepton production via decays $\phi\to e^+e^-$\, and $\mu^+\, \mu^-$\,. The comparison of
$\phi$ meson yields in the hadronic and electromagnetic
channels~\cite{NA49-Friese97,NA50-Jouan08,NA60-Floris08} can reveal interesting information about
the later stage of the collision~\cite{Johnson01,Filip01,Kolom02,Bleicher03}.

According to the SU(3) quark model the dominant component constituting the $\phi$ meson is a
spin-one bound state of $s$ and $\bar{s}$ quarks. Hence, the hadronic interactions of phi mesons
are subject to Okubo-Zweig-Iizuka rule, which tells that the interactions of a pure $(\bar{s}
s)$ state with non-strange hadrons are suppressed. In practice, the strict implementation of this
rule would imply vanishing of $\phi\, N N$ and $\phi \pi\rho$ couplings. Indeed, the
OZI-forbidden reactions are typically orders of magnitude smaller than the OZI-allowed, e.g.,
the ratio of $\omega$ to $\phi$ meson production cross section is about~\cite{Sibirtsev06}
\be
\frac{\sigma({\pi N\to \omega N})}{\sigma({\pi N\to \phi N})}\sim 75\,. \label{OZIsupp}
 \ee
This value is, however, three times smaller than the one which follows from the SU(3) symmetry and the
experimentally known deviation from the ideal mixing between singlet and octet vector
mesons~\cite{Lipkin76}. The phenomenological models are able to reproduce the experimental cross
section of the $\pi N\to \phi N$ reaction by taking into account the OZI-violating $\phi\rho\pi$
coupling adjusted to reproduce the $\phi\to\rho\pi$ decay width~\cite{Chung97,Titov00}. The role of
baryon resonances in this process was studied in ref.~\cite{Titov02,Doring08}.

Since the OZI suppression weakens the $\phi$ production only by the ordinary hadronic matter and
is lifted in the quark-gluon medium, the strong, of the order indicated in (\ref{OZIsupp}), enhancement of
the $\phi$ yield was proposed in~\cite{Shor85} as a signal of the quark-gluon plasma formation. An
enhancement of the $\phi$ yield was indeed observed in experiments albeit to a lesser degree: the
ratio of the $\phi$ yield in a nucleus-nucleus collision normalized by the number of participating
nucleon pairs to the $\phi$ yield in a proton-proton collision lies between 3 and 4 for AGS/SPS
energies~\cite{AGS-Back04,NA49-Alt08}. Within the hadrochemical model~\cite{Ko91} the factor 3 of
enhancement in $\phi$ production could be explained only by assuming a decrease of hadron masses
in hot and dense hadronic matter. In~\cite{Ko91} the main contribution to the $\phi$ yields are given by the
OZI-allowed process with {\it strangeness coalescence} $K\bar K\to \phi\rho$ and $K\Lambda\to
\phi N$\,. The string-hadronic transport model UrQMD~\cite{Bleicher99}, in which the $\phi$ is
produced mainly via $K\bar{K}\to \phi$ reactions and no in-medium effects are included,
systematically underestimates the $\phi$ production at SPS energies, cf. figure~9 in
ref.~\cite{NA49-Alt08}. The experiments suggest some similarity in production mechanism of $\phi$
mesons and kaons. In Au+Au collisions at 11.7$A$~GeV/c~\cite{AGS-Back04} and in Pb+Pb collisions at
158$A$~GeV/c~\cite{NA49-Afanasiev00,NA50-Alessandro03} it was observed that the $\phi$ yield grows
with the centrality more strongly than linearly in a similar way as the $K^+$ and $K^-$ yields do.
In contrast, the $\rho+\omega$ production scales linearly with a number of participating
nucleons~\cite{NA50-Alessandro03}.

At the SIS the $\phi$ production was studied at the beam energies about 2 GeV per nucleon, which
are less than the threshold energy of $\phi$ production in the nucleon-nucleon collision 2.6 GeV.
The FOPI collaboration measured $\phi$s in Ni+Ni reaction at 1.93
$A$GeV~\cite{FOPI-Mangiarotti03}. The reported ratio of $\phi$ to $K^-$ multiplicites is
$\langle\phi\rangle/\langle K^-\rangle\approx(0.44\pm0.16\pm 0.22)$ or $(1.7\pm 0.6\pm 0.85)$
depending on the temperature of the emitting source (130 MeV and 70 MeV, respectively) which has
to be assumed for an extrapolation to a full solid angle. These results are considerably larger
(at least by factor 4) than the very preliminary estimates announced in~\cite{Herrmann96}. Such a
large $\phi$ abundance cannot be described by the transport model~\cite{Chung97-2} where $\phi$s
are produced in reactions $BB\to BB\phi$ and $\pi B\to \phi B$ ($B= N,\,\Delta$) with the dominant
contribution from pion-nucleon reactions. The calculations with the IQMD code~\cite{Kampfer02}
underestimate the $\phi$ production data~\cite{FOPI-Mangiarotti03}, too. Even the inclusion of
other resonance channels $\rho B\to \phi N$ and $\pi N(1520)\to \phi N$ studied in
ref.~\cite{Barz02} would not fully accommodate the results~\cite{FOPI-Mangiarotti03} on $\phi$
multiplicites. Note that the strangeness coalescence process could not contribute much to the
$\phi$ yield at these energies since kaons have a long mean free path and most likely leave the
fireball right after its creation without any further interaction.

The large value of $\langle\phi\rangle/\langle K^-\rangle$ reported in~\cite{FOPI-Mangiarotti03}
seems to be confirmed by further analyses of the data~\cite{Herrmann08}. The preliminary results
by HADES experiments with Ar+KCl collision at 1.756 $A$GeV give also a comparable value for
$\langle\phi\rangle/\langle K^-\rangle\simeq 0.37\pm 0.13$~\cite{Agakichev09}.

In this paper we would like to address some puzzles of the $\phi$ production in heavy-ion
collisions at various energies and propose a new type of the $\phi$ production mechanism---{\it
the catalytic $\phi$ production} by strange particles, e.g. the reactions
\be
\pi Y\to \phi Y\,, \qquad \bar{K} N\to \phi Y\,, \qquad Y=\Lambda\,,\, \Sigma\,.
\label{Cathprod}
\ee
In contrast to the strangeness coalescence reaction, here the strangeness does not hide inside the
$\phi$ but stays in the system and the presence of $K$ mesons is unnecessary. The efficiency
of these reactions should be compared with the process $\pi N\to \phi N$, which is found to be
dominating in~\cite{Chung97-2}. The reactions (\ref{Cathprod}) are OZI allowed, so we win in cross
sections compared to $\pi N\to \phi N$ a factor of the order of 75, cf. (\ref{OZIsupp}). We lose, however,
in the smaller concentration of hyperons and anti-kaons compared to nucleons and pions,
respectively.

\begin{table}
\caption{Relative multiplicities of strange particles observed in nucleus collisions
at different colliding energies and the estimate of the lowest values of $\pi Y\to \phi Y$ and
$\bar K N\to \phi N$ cross sections at which this $\phi$ meson production mechanism could
dominate the $\pi N \to \phi N$ reaction ($\sigma(\pi N \to \phi N)\simeq 0.03$~mb).
The empirical date are taken from~\cite{Andronic06}.}
\label{tab:mult}
\begin{indented}\item[]
\begin{tabular}{lccccc}
\br
$E_{\rm lab}$                  &  2~{\rm GeV}/$A$  &  4~{\rm GeV}/$A$  &  6~{\rm GeV}/$A$ & 8~{\rm GeV}/$A$ & 10~{\rm GeV}/$A$ \\
\mr
$\frac{\langle K^+ \rangle}{N_{\rm part}}$    & $1.1\times 10^{-3}$ & $7.0\times 10^{-3}$ & $14\times 10^{-3}$ & $23\times 10^{-3}$ & $32\times 10^{-3}$\\
$\frac{\langle \Lambda \rangle}{N_{\rm part}}$& $1.3\times 10^{-3}$ & $8.4\times 10^{-3}$ & $15\times 10^{-3}$ & $20\times 10^{-3}$ & $30\times 10^{-3}$\\
$\frac{2\,\langle K^- \rangle}{\langle \pi^++\pi^-\rangle}$& ---  & $5.8\times 10^{-3}$ & $14\times 10^{-3}$ & $22\times 10^{-3}$   & $34\times 10^{-3}$\\
\mr
$\sigma(\pi \Lambda\to \phi Y)\gsim$         & 23~mb  & 3.5~mb &  2~mb & 1.5~mb & 1~mb\\
$\sigma(\bar{K} N\to \phi Y)\gsim $           & ---  & 5.2~mb &  2~mb & 1.3~mb & 0.9~mb\\
\br
\end{tabular}
\end{indented}
\end{table}

There is no experimental information about the cross sections for the reactions (\ref{Cathprod}),
therefore we will have to rely on some modeling. Let us first estimate how large the cross
sections for these reactions should be so that their contributions become comparable to $\pi N\to
\phi N$? This can be deduced from the following relations
\be
\sigma(\pi \Lambda\to \phi Y)\gsim \frac{N_{\rm part}}{\langle\Lambda \rangle}\,
\sigma(\pi N\to \phi N)\,,
\nonumber\\
\sigma(\bar{K} N\to \phi Y)\gsim \frac{\langle\pi^++\pi^-\rangle}{2\,\langle K^-\rangle}\, \sigma(\pi N\to \phi N)\,.
\nonumber
\ee
Here $N_{\rm part}$ number of nucleons participating in the collision, $\langle\Lambda \rangle$\,,
$\langle K^- \rangle$\,, $\langle\pi^+\rangle$\,, and $\langle\pi^-\rangle$  are the
multiplicities of the corresponding particles produced in the collisions. Using mid-rapidity data
for collisions at AGS energies, cf.~\cite{Andronic06}, listed in table~\ref{tab:mult} and the
experimental value $\sigma(\pi N\to \phi N)\simeq 0.03$~mb we get an estimate for the lower bound
of the $\pi \Lambda\to \phi Y$ and $\bar{K} N\to \phi Y$ cross sections. We see that a few mb
is enough to make the cathalitic $\phi$ meson production comparable with the conventional
mechanisms at AGS energies. For SPS energies these bounds could be even lower, since there are
more kaons and hyperons.

Although encouraging these rough estimates are not convincing yet. Indeed, the important is not
only how much strangeness is produced in the collision but also when it is produced and at what
temperature. The catalytic reactions (\ref{Cathprod}) will be operative if there is enough
strangeness produced and the medium is still hot enough. To study various scenarios quantitatively,
in section 2 we estimate the cross section of the reactions (\ref{Cathprod}) within a simple model.
In section 3 we set up a hadrochemical model for strangeness production. The $\phi$ production is
discussed in section 4.

\section{Estimate of $\pi\,Y\to \phi \, Y $  and
$KN\to \phi Y$ cross sections}

We consider the Lagrangian of lightest strange particles, i.e. hyperons and kaons, interacting
with pions and nucleons and include the $\phi$ meson as a gauge boson of the local U(1) symmetry
associated with the strangeness conservation
\be\nonumber
\mathcal{L}&=& \bar{\Lambda}\, \big(i\,\prt\cdot \gamma +\phi\cdot \gamma-m_\Lambda\big) \Lambda
+ \bar{\Sigma}_a\, \big(i\,\prt\cdot \gamma +\phi\cdot \gamma-m_\Sigma\big) \Sigma^a
\nonumber\\
&+& \case12\, \big(\prt-i\, \phi\big)_\mu\, K^\dag \, \big(\prt+i\, \phi\big)^\mu\,
K-\case12\, m_K^2\,K^\dag\, K
\nonumber\\
&+& C_{\pi\Lambda\Sigma}\,\big(\bar{\Lambda}\,\gamma_\mu\,\gamma_5\, \Sigma^a +
\bar{\Sigma}^a\,\gamma_\mu\,\gamma_5\,\Lambda  \big)\,\prt^\mu \pi_a
+ C_{\pi\Sigma\Sigma}\, i\,\epsilon_{abc} \big( \bar{\Sigma}^a\gamma_\mu\,\gamma_5
\Sigma^b \big) \prt^\mu \pi^c\,
\nonumber\\
&+& C_{KN\Sigma}\, \Big(
\bar{N}\, \gamma_\mu\, \gamma_5\, \tau_a\Sigma^a\, \big(\prt^\mu+i\, g_\phi\, \phi^\mu\big)\,
K+ \big(\prt^\mu-i\, g_\phi\, \phi^\mu\big)\, K^\dag\,\bar{\Sigma}_a\tau^a\, \gamma_\mu\, \gamma_5\, N\,
\Big)
\nonumber\\
&+& C_{KN\Lambda}\, \Big(
\bar{N}\, \gamma_\mu\, \gamma_5\, \Lambda\, \big(\prt^\mu+i\, g_\phi\, \phi^\mu\big)\,
K+ \big(\prt^\mu-i\, g_\phi\, \phi^\mu\big)\, K^\dag\,\bar{\Lambda}\, \gamma_\mu\, \gamma_5\, N\,
\Big)\,,
\label{lag}
\ee
where $\Sigma^a$ and $\pi^a$ denote the isospin triplets of $\Sigma$ baryons and pions; kaon and
nucleon isospin doublets are $K=(K^+, K^0)$ and $N=(p,n)$. The coupling $\phi K\bar{K}$ is fixed
from the $\phi\to K\bar{K}$ decay, $g_\phi=4.64$. The couplings
$C_{\pi\Lambda\Sigma}=0.63\,m^{-1}_\pi$, $C_{\pi\Sigma\Sigma}=-0.56\,m^{-1}_\pi$,
$C_{KN\Lambda}=-0.7\, m^{-1}_\pi$ and $C_{KN\Sigma}=0.4\, m^{-1}_\pi$ we take from the analysis
of meson baryon interactions~\cite{lk02}.

\subsection{$\pi\,Y\to \phi \, Y $ reaction}

\begin{figure}
\centerline{\includegraphics[width=9cm]{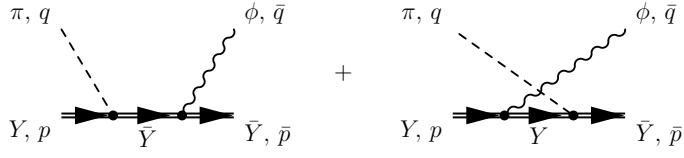}} \caption{Diagrams contributing
to a $\pi\,Y\to \phi \, Y $ reaction }
\label{fig:diag}
\end{figure}

In figure~\ref{fig:diag} the lowest-order  diagrams contributing to the  $\pi\,Y\to \phi \, Y $ process
are shown. The corresponding amplitude reads
\be
T&=& g_\phi\,C_{\pi\bar{Y}Y}\, i\,\bar{u}_{\bar{Y}}(\bar{p};\bar{s})\,\gamma^\mu\,
S_{\bar{Y}}(\bar p+\bar q)
\,(q\cdot\gamma) \,\gamma_5 u_{Y}(p;s)\,\epsilon^*_\mu(\bar{q};\lambda)
\nonumber\\
&+&
 g_\phi\,C_{\pi\bar{Y}Y}\,i\,\bar{u}_{\bar{Y}}(\bar{p};\bar{s})\,
(q\cdot \gamma)\,\gamma_5\,S_{Y}(p-\bar{q})
\,\gamma^\mu u_{Y}(p;s)\,\epsilon^*_\mu(\bar{q};\lambda)
\nonumber\\
&=& g_\phi\,C_{\pi\bar{Y}Y}\,\bar{u}_{\bar{Y}}(\bar{p};\bar{s})\, J^\mu_{\bar{Y}Y}(q)\,
u_{Y}(p;s)\,\epsilon^*_\mu(\bar{q};\lambda)\,,
\nonumber\\
&&
S_{Y}(p)=\frac{p\cdot \gamma+M}{p^2-M^2}\,,\quad
S_{\bar{Y}}(p)=\frac{p\cdot \gamma+\bar{M}}{p^2-\bar{M}^2}\,.
\label{ampl}
\ee
Here we use the following notations: $p$ and $q$ are the 4-momenta of a hyperon and a pion,
respectively, in the initial state, $\bar p$ and $\bar q$ are the 4-momenta of a hyperon and a phi meson in
the final state; $M$ and $\bar M$ are masses of the initial and final hyperons; $u_Y(p,s)$ denotes
the Dirac bispinor of a hyperon $Y$ with the momentum $p$ and the spin projection $s$;
$\epsilon^*_\mu(\bar{q};\lambda)$ is the wave function of an outgoing $\phi$ meson with
the polarization $\lambda$. For the
moment we do not specify in (\ref{ampl}) the isospin coefficient for the particular reaction.
Taking into account the on-shell relations for hyperons $(p\cdot \gamma-M)\, u(p,s)=0=\bar
u(\bar{p},\bar{s})\,(\bar{p}\cdot\gamma-\bar{M})$ we can reduce the current  $J^\mu_{\bar{Y}Y}(q)$
introduced in (\ref{ampl}) to the on-shell equivalent form
\be\nonumber
J_{\bar{Y}Y}^\mu(q)=i\,(M+\bar{M})\,\left\{ \gamma^\mu\,
S_{\bar{Y}}(\bar p+\bar q) \,\gamma_5+ \gamma_5\,S_{Y}(p-\bar{q})\,\gamma^\mu\right\}\,.
\ee
The current has the convenient properties
$\gamma_0\,\Big[J_{\bar{Y}Y}^\mu(q)\Big]^\dag\gamma_0 =J_{Y\bar{Y}}^\mu(-q)$
and $J^\mu_{if}(q)\, q_\mu=0$\,.
The cross section is determined by the amplitude squared, summed over the spin and
polarization of the final hyperon ($\bar s$) and phi meson ($\lambda$), and averaged over
the spin of the initial hyperon ($s$)
\be
\mathcal{F}_{\bar{Y}Y}&=&\case12\,\sum_{s,\bar{s},\lambda} \,T\,T^\dag \nonumber\\ &=&-
g^2_\phi\,C^2_{\pi\bar{Y}Y}\,\case12\,{\rm Tr}\left\{
(\bar{p}\cdot\gamma +\bar{M})
\,J_{\mu,\bar{Y}Y}(q)\,(p\cdot \gamma + M)\, J^\mu_{Y\bar{Y}}(-q)\right\}
\nonumber\\
&=&\phantom{-}
g^2_\phi\,C^2_{\pi\bar{Y}Y}\,(M+\bar{M})^2\,
\big[f_{\bar{Y}Y}(\bar{p},\bar{q},p)+g_{\bar{Y}Y}(\bar{p},\bar{q},p)
     +g_{Y\bar{Y}}(p,-\bar{q},\bar{p})\big]\,,
 \nonumber \ee
where
\be\fl
f_{\bar{Y}Y}(\bar{p},\bar{q},p) &=& \frac12\,{\rm Tr}\left\{
(\bar{p}\cdot\gamma +\bar{M})\,\gamma^\mu\,
S_{\bar{Y}}(\bar p+\bar q)  \,\gamma_5\,(p\cdot \gamma + M)\,
\gamma^\mu\, S_{Y}(p-\bar{q}) \,\gamma_5\right\}
\nonumber\\
\fl
&+& \frac12\,{\rm Tr}\left\{
(\bar{p}\cdot\gamma +\bar{M})\, \gamma_5\,S_{Y}(p-\bar{q})\,\gamma^\mu\,
(p\cdot \gamma + M)\, \gamma_5\,S_{\bar{Y}}(\bar p+\bar q) \,\gamma^\mu\right\}
\nonumber\\
\fl
&=&\frac{8}{((\bar{p}+\bar{q})^2-\bar{M}^2)\,((p-\bar{q})^2-M^2)}\,\Big[
\bar{M}\, M\,(2\, \bar{p}\cdot p+m_\phi^2)
\nonumber\\
\fl
&&
+(M-\bar{M})\, ( M\, \bar{p}\cdot \bar{q}+ \bar{M}\, p\cdot \bar{q})
-2\bar{p}\cdot(p-\bar{q})\, p\cdot(\bar{p}+\bar{q})\Big]\,,
\nonumber\\
\fl
g_{\bar{Y}Y}(\bar{p},\bar{q},p) &=& \frac12 \, {\rm Tr}\left\{
(\bar{p}\cdot\gamma +\bar{M})\,\gamma^\mu\,
S_{\bar{Y}}(\bar p+\bar q)  \,\gamma_5\,(p\cdot \gamma + M)\,
\gamma_5\,S_{\bar{Y}}(\bar p+\bar q) \,\gamma^\mu\right\}
\nonumber\\
\fl
&=&
\frac{4}{\big((\bar{p}+\bar{q})^2-\bar{M}^2\big)^2}\Big[
2\,(\bar{p}\cdot \bar{q}-\bar{M}^2)\,(p\cdot \bar{q}+ \bar{M}\, M)
\nonumber\\
\fl
&&- (\bar{p}\cdot p-2\,\bar{M}\, M)\,(2\,\bar{M}^2+m_\phi^2)\Big]\,.
\nonumber
\ee
Using the relations $2\, \bar{p}\cdot \bar{q} =s-\bar{M}^2-m_\phi^2$\,,
$2\,p\cdot \bar{q} =M^2+m_\phi^2-u$\,,
$2\,\bar{p}\cdot p=\bar{M}^2+M^2-t$\,, and $u+s+t=\bar{M}^2+M^2+m_\pi^2+m_\phi^2$
we can express $\mathcal{F}_{\bar{Y}Y}$ as a function of two invariants $s$ and $u$\,.
Then the total cross section can be evaluated as (cf.~\cite{LL})
\be
\sigma_{\bar{Y}Y}(s)=\frac{1}{32\,\pi\,s}\,\frac{p_{\phi\bar Y}(s)}{{p}_{\pi Y}(s)}\,
\intop_{-1}^{1} {\rm d} x\, \mathcal{F}_{\bar{Y}Y}(s,u(s,x))
\label{crosssec}
\ee
where
\be\fl
u(s,x)=\case12\, \Big(m^2_\phi+m_\pi^2+\bar{M}^2+M^2-s\Big)+
\frac{(M^2-m_\pi^2)\,(\bar{M}^2-m_\phi^2)}{2\, s}- 2\,p_{\pi Y}(s)\,p_{\phi \bar Y}(s)\, x\,,
\nonumber\\
\fl
p_{\pi Y}^2(s)=\case{1}{4s}\big((s-m_\pi^2-M^2)^2-4\, m_\pi^2\, M^2\big)\,,
\,\,
p_{\phi \bar Y}^2(s)=\case{1}{4s}\big((s-m_\phi^2-\bar{M}^2)^2-4\, m_\phi^2\, \bar{M}^2\big)\,.
\nonumber\ee

In calculations of the particle production in heavy-ion collisions
one usually uses isospin-averaged cross sections which can be defined as the sum
of cross sections for active (allowed) isospin channels divided by the total number of isospin
combinations of colliding particles. In $\pi\Lambda\to \phi \Sigma $ channel the following
three reactions are possible $\pi^{\pm,0} \Lambda\to \phi \Sigma^{\pm,0}$\,.
Since all these processes have the same cross section with the isospin coefficient equal to one,
the isospin-averaged cross section is
\be
\sigma_{\pi\Lambda\to \phi \Sigma}(s)=\sigma_{\Lambda\Sigma}(s)\,.
\label{sigLS}
\ee
In the $\pi\Sigma\to\phi \Sigma$ channel we have six possible reactions
$\pi^\pm\Sigma^\mp\to \phi \Sigma^0$\,,
$\pi^\pm\Sigma^0\to \phi \Sigma^\pm$\,, and
$\pi^0\Sigma^\pm\to \phi \Sigma^\pm $\,.
The entrance channel, $\pi\Sigma$, counts 9 possible isospin combinations.
The isospin averaged cross section is then
\be
\sigma_{\pi\Sigma\to\phi\Sigma}(s)=\frac23\,\sigma_{\pi^+\Sigma^-\to \phi \Sigma^0}(s)
=\frac23\,\sigma_{\Sigma\Sigma}(s)\,.
\label{sigSS}
\ee
The $\pi\Sigma\to\phi \Lambda$ channels is represented by three
reactions $\pi^\pm\Sigma^\mp\to \phi \Lambda$ and $\pi^0\Sigma^0\to \phi \Lambda$\,
out of nine available for the $\pi\Sigma$ entrance channel.
\be
\sigma_{\pi\Sigma\to\phi\Lambda}(s)=\frac13\,\sigma_{\pi^+\Sigma^-\to\phi\Lambda}(s)
=\frac13\sigma_{\Sigma\Lambda}(s)\,.
\label{sigSL}
\ee
\begin{figure}[t]
\centerline{\includegraphics[width=7cm]{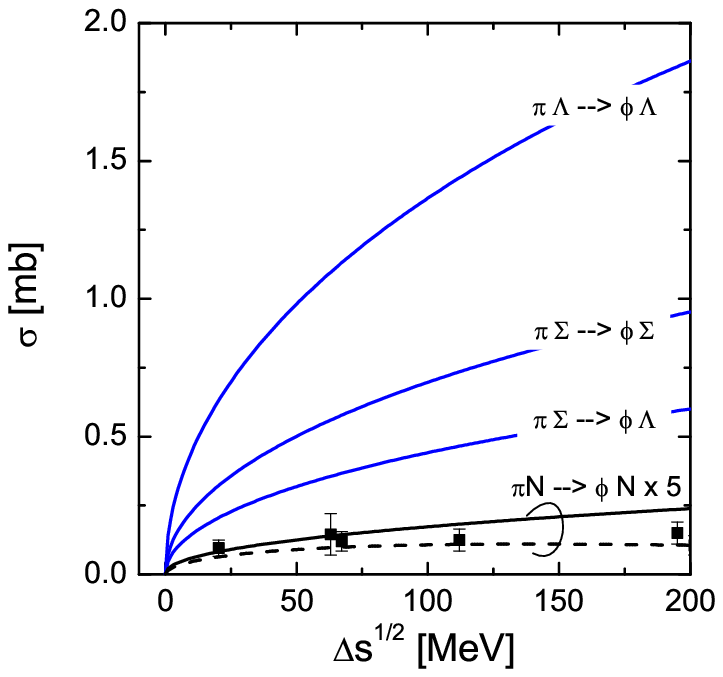}
\quad \includegraphics[width=7cm]{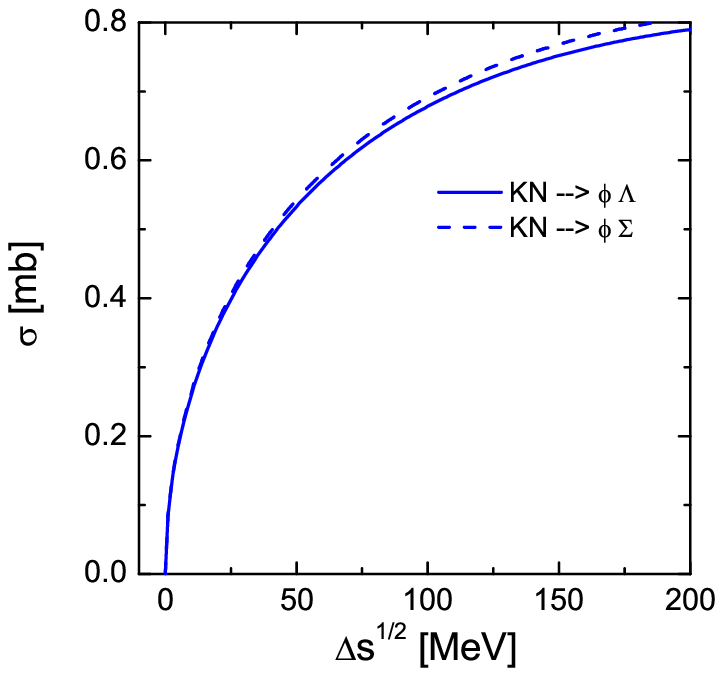}}
\caption{Isospin averaged cross sections of the $\phi$ meson production in $\pi Y\to \phi Y$  (left
panel) and in $KN\to \phi Y$ (right panel) reactions. On the left panel two lowest lines depict the
parameterisations of the $\pi^- p\to \phi n$ reaction from ref.~\cite{Kampfer02} (solid line)
and ref.~\cite{Sibirtsev97} (dashed line) in comparison with the experimental data from~\cite{Dahl67,Courant77}.
Both the paremeterizations and  the data are scaled up by factor 5.}
\label{fig:cross}
\end{figure}

The isospin-averaged cross sections (\ref{sigLS},\ref{sigSS},\ref{sigSL})
are shown in figure~\ref{fig:cross} as functions of the collision energy above the threshold
$\Delta s^{1/2}=s^{1/2}-s^{1/2}_{\rm th}$\,. For comparison, we
depict the cross section for the $\pi^- p \to \phi n $ reaction in the
parameterisation~\cite{Kampfer02} (solid line) and~\cite{Sibirtsev97} (dashed lines)
scaled up by factor 5. We see that the cross sections of the phi production on hyperons are
in average about 50 times larger than the production on nucleons (at $\Delta s^{1/2}\sim
100$~MeV)\footnote{Note that the isospin average cross section of $\pi N\to \phi N$ reaction is
$\sigma(\pi N\to \phi N)=\case12\, \sigma(\pi^- p\to \phi n)$.}.

\subsection{$\bar K N\to \phi Y$ reaction}
\begin{figure}
\centerline{\includegraphics[width=12cm]{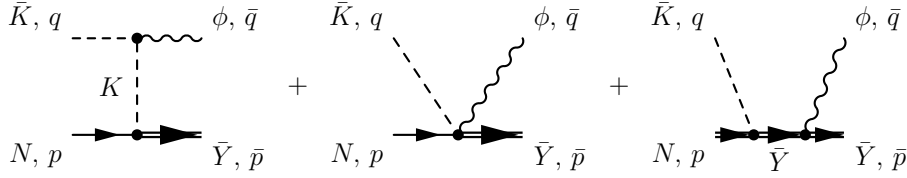}}
\caption{Diagrams contributing to a $\bar{K} N\to \phi \, Y $ reaction }
\label{fig:diagKN}
\end{figure}

The diagrams contributing to $\bar K N\to \phi Y$ processes in the lowest order
are depicted in figure~\ref{fig:diagKN}.
The reaction amplitude is
\be\fl
i\,T_{\bar{Y}N}&=& g_\phi\, C_{KN\bar{Y}}\,
\bar{u}_{\bar Y}(\bar p;\bar s)\,\Big( \gamma^\mu-(q-\bar{q})\cdot \gamma\,
D_K(q-\bar q)\,(2\,q-\bar q)^\mu\Big)\,\gamma_5\, u_N(p;s)\,\epsilon^*_\mu(\bar{q};\lambda)
\nonumber\\ \fl
&-&g_\phi\, C_{KN\bar{Y}}\, \bar{u}_{\bar Y}(\bar p;\bar s)\,
\gamma^\mu\,S_{\bar Y}(\bar p+\bar q)\,(q\cdot \gamma)\,\gamma_5\, u_N(p;s)\, \epsilon^*_\mu(\bar{q};\lambda)
\nonumber\\ \fl
&=&g_\phi\, C_{KN\bar{Y}}\, \bar{u}_{\bar Y}(\bar p;\bar s)\, J_{\bar{Y} N}(\bar q, q)\,
u_N(p;s)\, \epsilon^*_\mu(\bar{q};\lambda)\,,
\nonumber\\ \fl &&\qquad
 D_K(q)=(q^2-m_K^2)^{-1}\,.
\label{amplKN}
\ee
As before, we do not write explicitly the isospin factors for each reaction. They will be
restored later.
Using the on-shell conditions we rewrite the current $J_{\bar{Y} N}$ in the on-shell equivalent
form
\be
J_{\bar{Y} N}^\mu(\bar q, q)=(\bar M+ M)\, \Big(
D_K(\bar q- q)\, (2\, q-\bar q)^\mu+\gamma^\mu\,
S_{\bar Y}(\bar p+\bar q) \Big)\, \gamma_5\,.
\nonumber
\ee
The squared and spin summed amplitude is then
\be\fl
\mathcal{F}_{\bar{Y}N}&=&\case12\,\sum_{s,\bar{s},\lambda} \,T_{\bar YN}\,T_{\bar YN}^\dag
=\frac12\,{\rm Tr}\left\{
(\bar{p}\cdot\gamma +\bar{M})\,\gamma_5\,(p\cdot \gamma + M_N)\,\gamma_5\right\}
\, D_K^2(q-\bar q)\, (2\, q-\bar q)^2
\nonumber\\ \fl
&+&\frac12\,{\rm Tr}\left\{
(\bar{p}\cdot\gamma +\bar{M})\,\gamma_5\,(p\cdot \gamma + M_N)\,
\gamma_5\, S_{Y}(p-\bar{q}) \,\gamma_\mu\right\}\, D_K(q-\bar q)\, (2\, q-\bar q)^\mu
\nonumber\\ \fl
&+&\frac12\,{\rm Tr}\left\{
(\bar{p}\cdot\gamma +\bar{M})\,\gamma_\mu\, S_{Y}(p-\bar{q}) \,\gamma_5\,(p\cdot \gamma + M_N)\,
\gamma_5 \right\}\, D_K(q-\bar q)\, (2\, q-\bar q)^\mu
\nonumber\\ \fl
&+&\frac12 \, {\rm Tr}\left\{
(\bar{p}\cdot\gamma +\bar{M})\,\gamma^\mu\,
S_{\bar{Y}}(\bar p+\bar q)  \,\gamma_5\,(p\cdot \gamma + M_N)\,
\gamma_5\,S_{\bar{Y}}(\bar p+\bar q) \,\gamma^\mu\right\}
\nonumber\\ \fl
&=& g_{\bar{Y} N}(\bar{p},\bar{q},p) - 2\,
\frac{(2\, q-\bar q)^2\,\big(\bar p \cdot p -\bar M\, M_N \big)}{\big((q-\bar q)^2-m_K^2\big)^2}
\nonumber\\ \fl
& - & 4\,\frac{
\Big( \big(\bar p \cdot p -\bar M\, M_N \big)\, (2\, q-\bar q)\cdot (2\,\bar p +\bar q)
+2\, (p\cdot\bar q)\, (\bar p\cdot q) -2\, (\bar p\cdot\bar q)\, (p\cdot q)
\Big)}{\big((q-\bar q)^2-m_K^2\big)\, \big((\bar p+\bar q)^2-\bar M^2\big)}
\nonumber
\ee
The function $g_{\bar{Y}N}$ follows from  $g_{\bar{Y}Y}$ after the replacement $M\to M_N$\,.
Using the kinematic relations
$2\, \bar p\cdot \bar q=s- \bar M^2-m_\phi^2$\,,
$2\,  p\cdot q=s- M_N^2-m_K^2$\,,
$2\, \bar p\cdot p= \bar M^2+M_N^2-t$\,,
$2\, \bar q\cdot q= m_\phi^2-m_\pi^2- t$\,,
$2\, \bar p\cdot q =s+t -M_N^2-m_\phi^2$\,, and
$2\, p\cdot \bar q =s+t-\bar M^2-m_\pi^2$ we can express $\mathcal{F}_{\bar{Y}N}$ as
a function of invariants $s$ and $t$\,.
The cross section of $\bar K N\to Y \phi$ is given by
\be
\sigma_{\bar{Y}N}(s)=\frac{1}{32\,\pi\,s}\,\frac{p_{\phi\bar Y}(s)}{p_{K N}(s)}\,
\intop_{-1}^{1} {\rm d} x\, \mathcal{F}_{\bar{Y}N}(s,t(s,x)) \,,
\label{sigKN}\\ \fl
t(s,x)=m_\phi^2+m_K^2-2\, \sqrt{m_\phi^2+p_{\phi\bar Y}^2(s)}\,
\sqrt{m_K^2+p_{KN}^2(s)}+2\,p_{K N}(s)\,p_{\phi\bar Y}(s)\, x \,,
\nonumber\\ \fl
p_{KN}^2(s)=\case{1}{4s}\big((s-m_K^2-M_N^2)^2-4\, m_K^2\, M_N^2\big)\,,\quad
p_{\phi \bar Y}^2(s)=\case{1}{4s}\big((s-m_\phi^2-\bar M^2)^2-4\, m_\phi^2\, \bar M^2\big)\,.
\nonumber
\ee
The isospin averaged cross section for the $KN\to \phi \Lambda$ reaction is equal to
\be
\sigma_{KN\to \phi \Lambda}(s)=\frac13\,\sigma_{\Lambda N}(s)
\label{sigNL}
\ee
where we use that $\sigma_{K^-p\to \phi \Lambda}(s)=\sigma_{\bar{K^0}n\to \phi\Lambda}(s)$\,.
Evaluating the isospin averaged cross section for the $KN\to \phi \Sigma$ reaction
we take into account that the reactions with charged $\Sigma$ hyperon in the final state
have additional isospin factor 2, then
\be
\sigma_{KN\to \phi \Sigma}(s)=\sigma_{\Sigma N}(s)\,.
\label{sigNS}
\ee
The cross sections (\ref{sigNL}) and (\ref{sigNS}) are shown on the right panel in
figure~\ref{fig:cross}. In comparison with the $\pi^-p\to \phi n$ cross section
the $KN\to \phi Y$ cross sections are about 60 times larger at $\Delta s^{1/2}\sim 100$~MeV.

\section{Strangeness production}

In order to estimate the efficiency of reactions with the $\phi$ production on hyperons and kaons,
we have to study, first,  the dynamics of the strangeness production in the course of a nuclear
collision. This question was addressed within hadro-chemical models
in~\cite{Ko88,Barz88,Barz90,Russkikh91,Russkikh92}. In our consideration we make the following
assumptions: (a) the fireball matter is baryon-dominated; (b) strangeness can be considered
perturbatively, i.e, the number of produced strange particles, say kaons, is much smaller than the
baryon number. In the baryon-dominated matter the particles carrying strange quarks, $K^-$\,,
$\bar{K}^0$\,, $\Lambda$\,, $\Sigma$ and heavier hyperons have short mean free path. Hence, we can
assume that they are confined inside the fireball and remain in thermal equilibrium with other
species till the fireball breaks up. Oppositely, the particles with anti-quark , $K^{+}$ and $K^0$
mesons, have larger mean free paths and can leave the fireball at some earlier stage of the
collision. Based on this observation the strangeness separation scenario was formulated in
refs.~\cite{Ko83,Kolomeitsev95,Tomasik05}. It can be suitable for nucleus-nucleus collisions at
AGS energies. The analysis in~\cite{Tomasik05} shows that it can be also partially valid at SPS
energies. However, at what stage of the collision kaons ($K^+$ and $K^0$) leave the fireball is not important
for our further calculations if the concentrations of strange particles remain small. Since
in such a case the reactions with strangeness annihilation,  $KY\to \pi N$ and   $K\bar K\to \pi \pi$,
can always be neglected and the number of strange particles accumulated in the system is equal to
the number of produced kaons.

The time dependence of the fireball temperature and baryon density will be parameterized in the
form of a scaling solution of the hydrodynamic equations~\cite{Bondorf78}
\be
T(t)=\frac{T_m}{(t^2/t^2_0+1)^\alpha}\,,\quad
\rho_B(t) &=&\frac{\rho_m}{(t^2/t^2_0+1)^{3\,\alpha/2}}
\,,\qquad
\label{Tr-evol}
\ee
where $T_m$ and $\rho_m$ are the initial (maximal) temperature and density of the fireball, and
$t_0$ is the typical time scale of the fireball expansion.

The evolution of the kaon density is described by the differential equation
\be
\frac{{\rm d} \rho_{K}}{{\rm d} t}-\rho_K(t)\, \frac{\dot\rho_B(t)}{\rho_B(t)}
&=&\mathcal{R}(t)\,,\qquad \rho_K(0)=0\,.
\label{Kprod}
\ee
The second term on the left-hand side takes into account the reduction of the kaon density due to
the expansion of the system (dilution). We exploit here the relation for the volume change
$\dot V(t)/V(t)=-{\dot\rho_B(t)}/{\rho_B(t)}$\,. The kaon production rate on the right-hand side
is determined by the processes with $\pi N$, $\pi \Delta$, $NN$, $\pi\pi$ and  $N\Delta$
in the initial states
\be
\mathcal{R} &=&\kappa_{\pi N}^{K X}\, \rho_{\pi}\, \rho_{N} +
\kappa_{\pi \Delta}^{KX}\, \rho_\pi\, \rho_\Delta +
\kappa_{NN}^{KX}\, \rho_N^2+
\kappa_{\pi\pi}^{K\bar K}\, \rho_\pi^2
+\kappa_{N\Delta}^{KX}\, \rho_N\, \rho_\Delta\,.
\label{Rate}
\ee
As argued above we do not include on the right hand side in (\ref{Kprod}) the rates of kaon
annihilation processes. They are clearly absent if $K$ mesons leave fireball immediately after their
creation. If $K$ mesons stay in the system the annihilation reactions are still very suppressed as
being proportional to $(\rho_K/\rho_B)^2\ll 1$\,.

The transport coefficient
$$\kappa_{ab}^{KX}=\frac{\langle \sigma_{ab}^{KX} v_{ab} \rangle}{1+\delta_{ab}}$$
is the cross section $\sigma_{ab}^{KX}(s)$ of the binary reaction $a+b\to K+X$
averaged with the particle relative velocity $v_{ab}$ over the momentum distributions of colliding
particles. Following ref.~\cite{Ko88} we write
\be\fl
\kappa_{ab}^{KX}(T)
\label{kappadif}
 =
\frac{\dsp
\int_{\sqrt{s}_{\rm th}}^{\infty} {\rm d} x\,
\sigma_{ab}^{KX}(x^2)\, K_1(\case{x}{T})\,
[x^2-(m_a+m_b)^2]\,[x^2-(m_a-m_b)^2]}
{(1+\delta_{ab})\,4\,m_a^2\, m_b^2\, T\, K_2(m_a/T)\, K_2(m_b/T)},
\ee
where $K_1(x)$ and $K_2(x)$ are the MacDonald's functions of the first and second orders,
and  $\sqrt{s}_{\rm th}$ is the reaction threshold.
Among the $\pi N$ induced reactions we include $\pi N\to K\Lambda$\,, $\pi N\to K\Sigma$
and $\pi N\to K\bar K N$ processes: $\kappa_{\pi N}^{K X}=\kappa_{\pi N}^{K \Lambda}
+\kappa_{\pi N}^{K \Sigma}+ \kappa_{\pi N}^{K \bar K \,N}$\,.
For the isospin averaging we use the relations~\cite{Sibirtsev97,Cugnon84}
\be\fl
\sigma(\pi N\to \Lambda K) &=& \frac12\, \sigma(\pi^- p\to \Lambda K^0)\,,
\nonumber \\ \fl
\sigma(\pi N\to \Sigma K) &=& \frac12\, \Big(
\sigma(\pi^- p\to \Sigma^- K^+)+
\sigma(\pi^- p\to \Sigma^0 K^0)+
\sigma(\pi^+ p\to \Sigma^+ K^+)\Big)\,,
\nonumber\\ \fl
\sigma(\pi N \to N K \bar K) &=& 3\, \sigma (\pi^-  p\to N K^0 K^-)\,.
\nonumber
\ee
The parameterizations for the elementary $\pi N\to K Y$ cross sections are taken from
refs.~\cite{Tsushima94, Tsushima97} and for the $\pi^-  p\to N K^0 K^-$ cross section
from ref.~\cite{Sibirtsev97}.
In the $\pi \Delta$ channel we include only the process with $\pi\Lambda$ and $\pi\Sigma$ in the
final state: $\kappa_{\pi\Delta}=\kappa_{\pi\Delta}^{K\Lambda}+\kappa_{\pi\Delta}^{K\Sigma}$\,.
The isospin averaged cross section of the  $\pi \Delta\to Y K$
reactions we derive  at hand of the relations (21-31) in ref.~\cite{Tsushima94}
\be\nonumber
\sigma(\pi\Delta\to \Sigma K) &=&
\frac{1}{36}\, \Big(5\, \sigma(\pi^-\Delta^{++}\to \Sigma^0 K^+) +
6\, \sigma(\pi^0\Delta^0\to \Sigma^- K^+)
\\ \nonumber
&& + 21\, \sigma(\pi^+\Delta^0\to \Sigma^0 K^+)
+ 8\, \sigma(\pi^+\Delta^-\to \Sigma^- K^+)
\Big)
\\ \nonumber
\sigma(\pi\Delta\to \Sigma K) &=&\frac13\,
\sigma(\pi^-\Delta^{++}\to \Lambda K)
\ee
with the elementary cross section given in ref.~\cite{Tsushima94}.
The transport coefficient of kaon production in the $N N$ channel
has three contributions $\kappa_{NN}^{K X}=  \kappa_{NN}^{K\Lambda N}+\kappa_{NN}^{K\Sigma N}+
\kappa_{NN}^{K\bar K N N}$ and the parameterisations of the elementary
$NN\to KNY$ and $NN\to K\bar K NN$ cross sections are taken from ref.~\cite{Cassing97}.
The cross section for the $\pi\pi\to K\bar K$ reaction which determines $\kappa_{\pi\pi}^{K\bar K}$
we take from ref.~\cite{Cassing97} too.
The cross sections for reactions $N\Delta\to KYN$ which contribute to
$\kappa_{N\Delta}^{KX}=\kappa_{N\Delta}^{KN\Lambda}+\kappa_{N\Delta}^{KN\Sigma}$
are taken from ref.~\cite{Tsushima99}

In our model consideration we neglect contributions from other process with kaon production
which have higher thresholds and, therefore, are less relevant in the range of temperatures
we consider below $T\lsim m_\pi$\,.

The interaction among pions, nucleons and deltas is strong enough to sustain
them in thermal and chemical equilibrium during the whole fireball evolution. Hence, their
densities are given by
\be
\rho_\pi(t) &=& \frac{3}{2\pi^2}\int_0^\infty {\rm d} p\ p^2\, \exp\Big({-\sqrt{m_\pi^2+p^2}/T(t)}\Big)\,,
\nonumber\\
\rho_N(t) &=&  \frac{2}{\pi^2}\int_0^\infty\, {\rm d} p\, p^2\,
\exp\Big({\mu_B(t)/T(t)-\sqrt{m_N^2+p^2}/T(t)}\Big)\,,
\nonumber\\
\rho_\Delta(t) &=& \frac{8}{\pi^2}\int_0^\infty\, {\rm d} p\, p^2\,
\exp\Big({\mu_B(t)/T(t)-\sqrt{m_\Delta^2+p^2}/T(t)}\Big)\,.
\ee
We disregard the finite width of $\Delta$s and treat them as stable particles with the
mass $m_\Delta=1232$~MeV.
Neglecting the contributions of hyperons, heavier resonances and
anti-particles we can find the baryon chemical potential from the following relation
\be\fl
\mu_B(t)=-T(t)\,\ln\left[
\frac{2}{\rho_B(t)}\int_0^\infty\, \frac{{\rm d} p\, p^2}{\pi^2}\,
\left(e^{-\sqrt{m_N^2+p^2}/T(t)} + 4\, e^{-\sqrt{m_\Delta^2+p^2}/T(t)}\right)
\right]\,.
\label{muB}
\ee
\begin{figure}
\centerline{
\parbox{7cm}{\includegraphics[width=7cm]{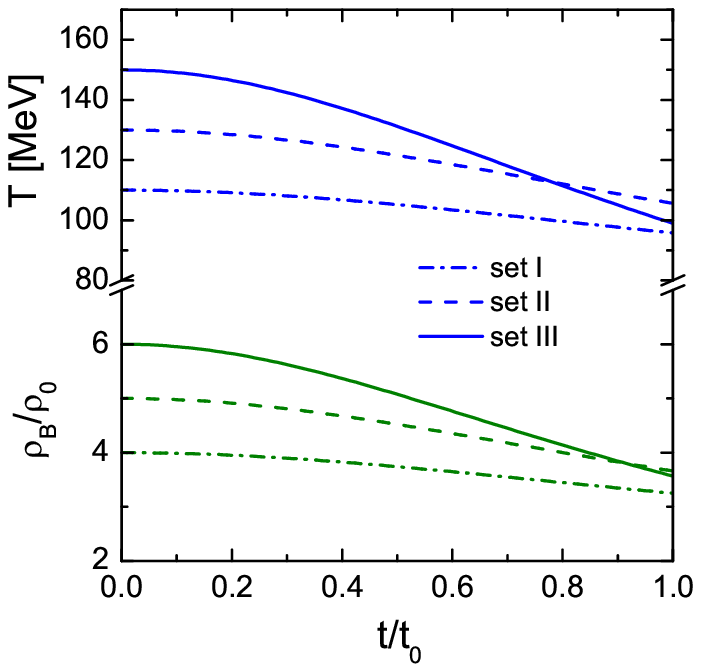}}
\quad
\parbox{7cm}{\includegraphics[width=7cm]{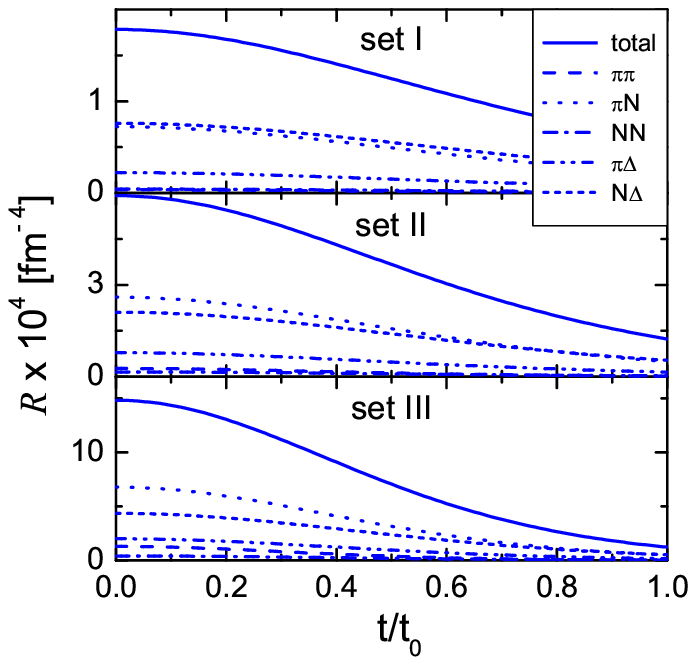}}
}
\caption{Left panel: temperature and baryon density as functions of time for three parameter sets
(\ref{cases}). Right panel: Rates of kaon production and contribution from different processes, cf. (\ref{Rate}),
respectively}
\label{fig:T-r}
\end{figure}
In order to study different fireball evolution scenarios we consider
three sets of parameters for the maximal temperature and density
\be
&{\rm I.}\quad  & T_m=110~{\rm MeV}\,,\quad \rho_m=4~\rho_0\,,\quad \alpha=0.2\,,
\nonumber\\
&{\rm II.}\quad &T_m=130~{\rm MeV}\,,\quad \rho_m=5~\rho_0\,, \quad \alpha=0.3\,,
\nonumber\\
&{\rm III.}\quad& T_m=150~{\rm MeV}\,,\quad \rho_m=6~\rho_0\,,\quad \alpha=0.5\,,
\label{cases}
\ee
where $\rho_0=0.17$\,fm$^{-3}$ is the nuclear saturation density.
The dependence of the temperature and density on time is exemplified on the left panel
in figure~\ref{fig:T-r}. On the right panel in figure~\ref{fig:T-r}
the rates of kaon productions in various processes are shown. The most efficient ones are processes
with $\pi N$ and $N\Delta$ in the incoming channel.

When all ingredients entering the rate (\ref{Rate}) are specified the density
of produced kaon follows after integrating (\ref{Kprod})
\be
\rho_K(t)=\rho_B(t)\, \intop_0^t\, {\rm d}t' \frac{\mathcal{R}(t')}{\rho_B(t')}
=\rho_B(t)\,t_0\, \intop_0^{t/t_0}\, {\rm d}x \frac{\mathcal{R}(T(t_0\, x),\rho_B(t_0\, x))}{\rho_B(t_0\,x)}
\,. \label{Kdens}
\ee The produced kaons escape from the fireball leaving behind a medium with some net strangeness
accumulated in form of $\bar K$, $\Lambda$ and $\Sigma$. ($\Xi$ and $\Omega$ can be neglected since
their abundances  are of the higher orders in terms of the small parameter $\rho_K/\rho_B$.) The occupation of
particles caring the strange quark is according to their statistical weights
\be
\rho_i(t) &=&\rho_K(t)\, \frac{n_{i}(t)}{\sum_{i=\bar{K},\Lambda,\Sigma} n_i(t)}\,,\quad
i=\bar{K}\,,\,\Lambda\,,\,\Sigma\,,
\nonumber\\
n_{\bar K}(t) &=&\phantom{3} \int_0^\infty \frac{{\rm d} p \, p^2}{\pi^2}\,
\exp\Big({-\sqrt{m_K^2+p^2}/T(t)}\Big)\,,
\nonumber\\
n_\Lambda(t) &=&\phantom{3} \int_0^\infty \frac{{\rm d} p \, p^2}{\pi^2}\,
\exp\Big({\mu_B(t)/T(t)-\sqrt{m_\Lambda^2+p^2}/T(t)}\Big)\,,
\nonumber\\
n_\Sigma(t) &=& 3\int_0^\infty \frac{{\rm d} p \, p^2}{\pi^2}
\exp\Big({\mu_B(t)/T(t)-\sqrt{m_\Sigma^2+p^2}/T(t)}\Big)\,.
\label{strange-dist}
\ee
We assume, thereby, that the reactions redistributing the strange quark among $\bar K$ and hyperons are
swift.

\begin{figure}
\centerline{
\parbox{14cm}{\includegraphics[width=14cm]{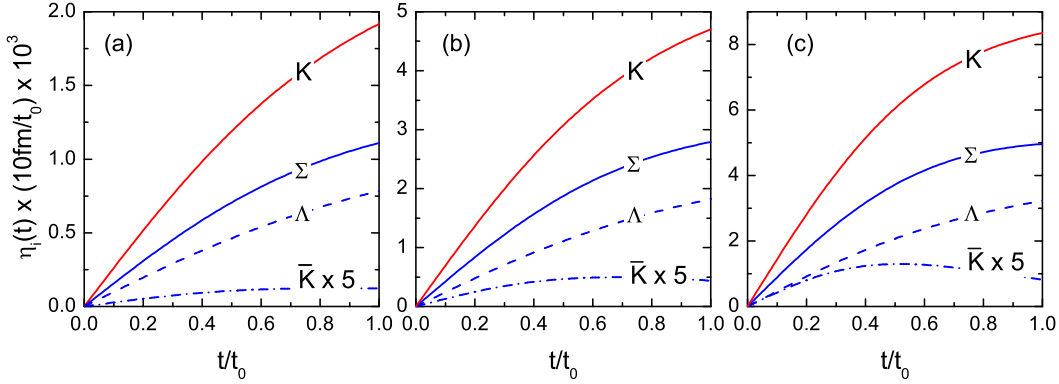}}
} \caption{Evolution of the number of kaons ($K$) produced in the course of the collision
and the number of strange particles ($\Sigma$, $\Lambda$ and $\bar K$) accumulated in the
fireball. Three panels (a-c) corresponds to the different sets of the maximal temperature and density
(I-III) in (\ref{cases}). } \label{fig:str}
\end{figure}

The evolution of strangeness  in the course of a collision is illustrated
in figure~\ref{fig:str} where we depict the ratios of strange particles, $K$\,, $\bar{K}$\,, $\Lambda$\,,
and $\Sigma$, to the baryon density
\be
\eta_i(t)=\rho_i(t)/\rho_B(t)\,.
\ee
Note that in view of eqs.~(\ref{Tr-evol}) and (\ref{strange-dist})
the kaon density (\ref{Kdens}) and the densities of other strange particles are linearly
proportional to the parameter $t_0$\, for the fixed $t/t_0$. Presenting the results in figure~\ref{fig:str} we
take into account this scaling, i.e., the lines are to be scaled by factor $t_0/(10~{\rm
fm})$.

We observe that strangeness is dominantly presented in form of $\Sigma$ hyperons
and that the number of produced strange particles doubles roughly with an increase of the
temperature by every 20 MeV from case I to case III in (\ref{cases}).

\section{$\Phi$ mesons in heavy-ion collisions}
In this section we discuss how the catalytic reactions can influence various aspects of $\phi$
production in heavy-ion collisions.

\subsection{$\Phi$ meson production}

\begin{figure}
\centerline{
\parbox{14cm}{\includegraphics[width=14cm]{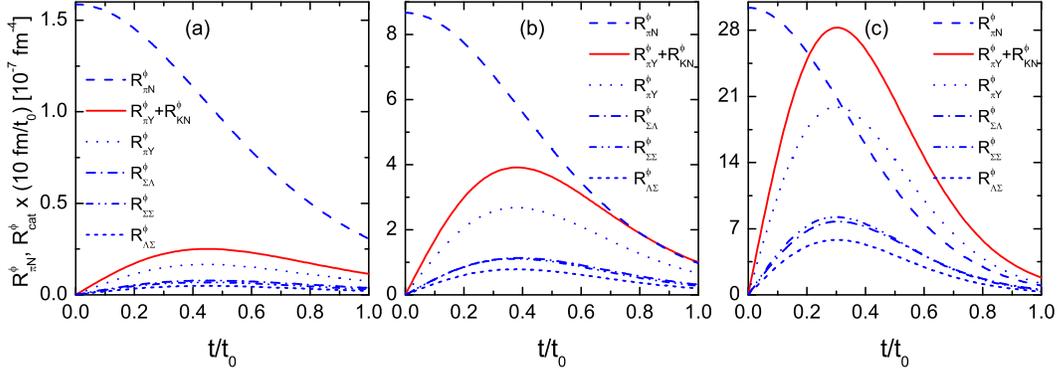}}
} \caption{Production rate of $\phi$ mesons in various catalytic reactions and their sum (solid
line) in comparison with the $\pi N\to \phi N$ reaction rate (dashed line). Different panels (a--c)
show the results for sets I--III, respectively} \label{fig:FRates}
\end{figure}

First we consider whether the catalytic reaction can be an efficient source of $\phi$ mesons compared
to a conventional reactions.
The rate equation for the $\phi$ meson production has the form similar to (\ref{Kprod})
\be\fl\nonumber
\frac{{\rm d} \rho_{\phi}}{{\rm d} t}-\rho_\phi(t)\, \frac{\dot\rho_B(t)}{\rho_B(t)}
= R^\phi_{\pi N}(t)+\sum_{\bar{Y},Y=\Lambda,\Sigma} R^\phi_{Y\bar{Y}}(t)+ R_{\bar K N}^\phi(t)+
\dots\,\, ,
\label{Phirates}
\ee
where on the right hand side we explicitly write down the $\phi$ production rates from
the conventional $\pi N\to \phi N$ reactions
$R^\phi_{\pi N}(t) = \kappa_{\pi N}^{\phi N}\, \rho_\pi\, \rho_N\,$
and the catalytic reactions on hyperons
$
R_{\pi Y}^\phi=\sum_{\bar{Y},Y=\Lambda,\Sigma} R^\phi_{Y\bar{Y}}\,,
\quad
R^\phi_{Y\bar{Y} }(t) = \kappa_{\pi Y}^{\phi \bar Y}\, \rho_\pi\, \rho_{Y}
$
and on anti-kaons
$
R^\phi_{\bar KN}(t) = \left(\kappa_{\bar{K} N}^{\phi \Lambda}+\kappa_{\bar{K} N}^{\phi \Sigma}\right)\,
\rho_{\bar K}\, \rho_{N}\,.
$
The ellipses stand for further $\phi$ production reactions like those discussed
in~\cite{Kampfer02} and the $\phi$ absorption processes, e.g., $\phi N\to K\, Y$ and $\phi Y\to
\pi Y$, etc. The rates of various process are shown figure~\ref{fig:FRates}. The $\phi$ production
in $\pi N$ collisions starts of course at the very beginning and gradually falls off as the fireball
cools down and expands. The rates of catalytic reactions increase, the more strange particles are
produced, reach the maximum at (0.3--0.4)$\,t_0$ and drop off later. For set~I with the maximal
temperature $T_m=110$~MeV the net rate of catalytic reactions yields to that of $\pi N\to \phi N$.
For set~II with $T_m=130$~MeV the rates become comparable for times $\gsim 0.6\, t_0$. For
$T_m=150$~MeV in set~III the catalytic production of $\phi$ exceed the rate of the $\pi N\to\phi
N$ reaction already at $0.2\, t_0$. Note that the numerical values for the rates in
figure~\ref{fig:FRates} correspond to the fireball expansion time $t_0=10$~fm. If the collision
lasts longer then the curves for catalytic reactions have to be scaled up by a factor
$t_0/(10~{\rm fm})$, since the number of the strange particles is proportional to the expansion
time. This will make the catalytic reaction efficient even for the smaller temperature in
set~I.

The above estimates show that the catalytic processes can be potentially important for the $\phi$
production in heavy-ion collisions. It is interesting now to investigate how they can affect other
$\phi$ production patterns, and whether their dominance can lead to any contradiction with
experimental data.

\subsection{Centrality dependence}
We discuss now the centrality dependence of the $\phi$ production.
Changing centrality one changes the volume of the
system. We will use the mean number of projectile participants, $N_{\rm pp}$, as the measure for initial
volume of the fireball created in the collision, $V\propto N_{\rm pp}$. The $N_{\rm pp}$ can be
directly related to the energy deposited in the zero-degree calorimeter, cf.~\cite{AGS-Back04}. If
there is only one changing parameter with the unit of length as in the case of a symmetrical collision at a
fixed collision energy, $l\sim V^{1/3}\propto
N_{\rm pp}^{1/3}$, the scaling properties of hydrodynamics imply that the collision time is of the
order $t_0\sim l/c\propto N_{\rm pp}^{1/3}$~\cite{Russkikh92}. The multiplicities of particle species
which reach the full chemical equilibrium, e.g. pions and $\Delta$s, are proportional to the volume
of the fireball and scale, therefore, as $N_\pi\propto V\propto N_{\rm pp}$ and $N_\Delta\sim N_{\rm
pp}$\,. The multiplicity of kaons ($K^+$), which have a long mean free path and leave the
system right after being produced, scales as $N_{K^+}\propto V\, t_0\propto N_{\rm pp}^{4/3}$\,.
According to (\ref{strange-dist}) other strange particles should follow the $K^+$ scaling,
$N_{\Lambda,\Sigma,\bar{K}}\sim N_{\rm pp}^{4/3}$\,. The number of produced $\phi$ mesons can be
estimated as
\be
N_\phi
&\sim & \kappa_{\pi N }\, \frac{N_\pi\, N_{\rm pp}}{V}\, t_0+\kappa_{\pi Y}\, \frac{N_\pi\, N_{Y}}{V}\, t_0
+\kappa_{\bar K N}\, \frac{N_{\bar K}\, N_{\rm pp}}{V}\, t_0
\nonumber\\
&\sim &
a_{\rm conv}\,N_{\rm pp}^{4/3} + a_{\rm cat} \,N_{\rm pp}^{5/3}
\nonumber
\ee
The term $\sim N^{4/3}_{\rm pp}$ is due to the conventional production reactions like $\pi N\to\phi
N$ whereas the term $\sim N^{5/3}_{\rm pp}$ one corresponds to the catalytic reactions which
contribute with an additional factor $t_0$\,. For the ratios experimentally observed in
ref.~\cite{AGS-Back04} we find
\be
\frac{N_\phi}{N_\pi}\sim a\, \Bigg(\frac{N_{\rm pp}}{A}\Bigg)^{1/3}+b\, \Bigg(\frac{N_{\rm pp}}{A}
\Bigg)^{2/3}\,,
\quad
\frac{N_\phi}{N_{K^+}}\sim a' + b'\, \Bigg(\frac{N_{\rm pp}}{A}\Bigg)^{1/3}\,,
\label{Phiratios}
\ee
\begin{figure}
\centerline{
\parbox{7cm}{\includegraphics[width=7cm]{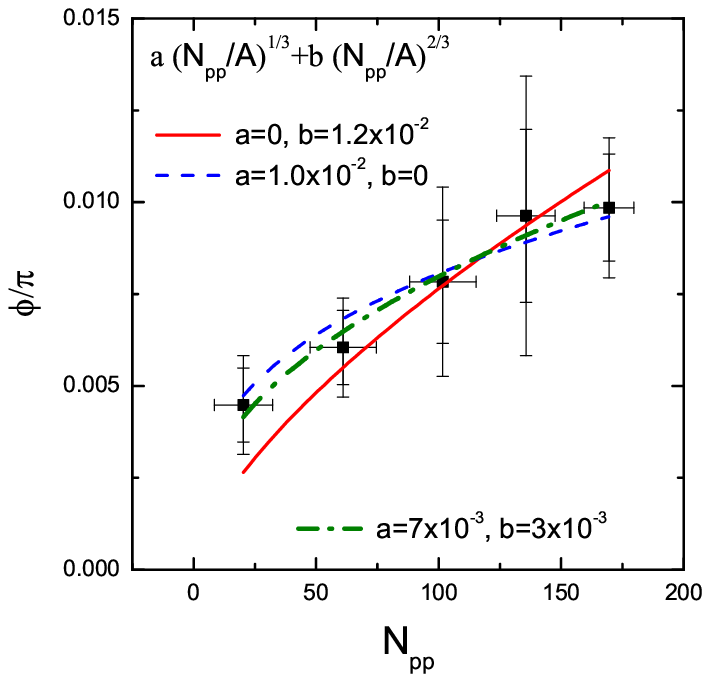}}
\quad
\parbox{7.1cm}{\includegraphics[width=7.1cm]{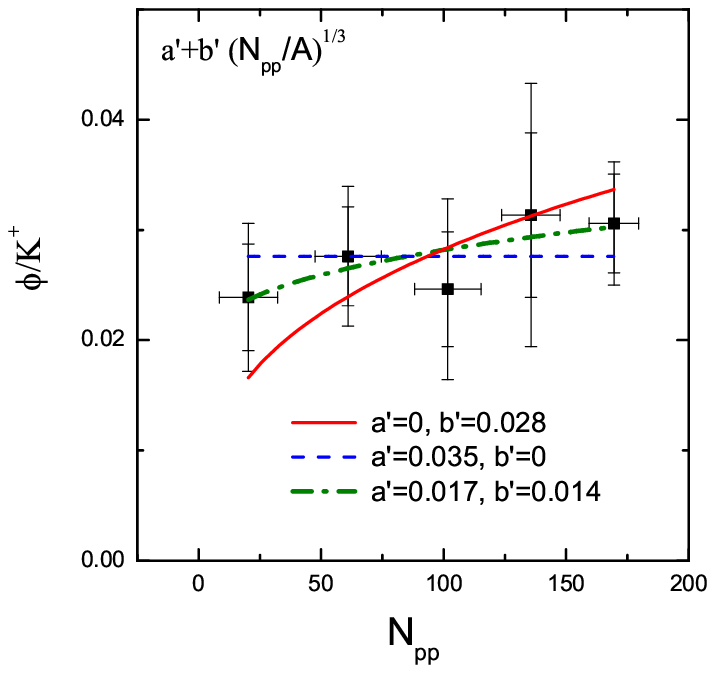}}
}
\caption{The $\phi/\pi$ (left panel) and $\phi/K^+$ (right pane) as a function of the mean number
of projectile participants. The data points from ref.~\cite{AGS-Back04} correspond to Au+Au
collisions at 11.7~$A$GeV. The curves show the relations~(\ref{Phiratios}). The value of
the parameters are given on the plots ($A=197$).}
\label{fig:Central}
\end{figure}
where $a\,,a'$ and $b\,,b'$ parameterize the relative strength of conventional and
catalytic processes, and $A$ is the number of nucleons in the colliding nuclei.
In figure~\ref{fig:Central} we compare these parameterizations with the only
available data of the $\phi$ yield centrality dependence for Au+Au collisions at
11.7~$A$GeV~\cite{AGS-Back04}. First, we adjust parameters $a,a'$ and $b,b'$
separately and obtain dashed and solid curves, respectively. Because of large error bars both fits
with $a,a'=0$ or $b,b'=0$ equally well go through data points. So the data on
centrality dependence cannot rule out even the dominance of the catalytic reactions. The optimal
fits are reached when both parameters are activated, dashed-dotted lines in
figure~\ref{fig:Central}. Comparing the values of parameters $a,a'$ and $b,b'$ from
these fits we can conclude that the catalytic reaction contribution can be about 30\%-40\%
for $N_{\rm pp}=A$.

\subsection{$\phi$ rapidity distribution}

The systematics of $\phi$ rapidity distributions in Pb+Pb collisions at the SPS is reported in
ref.~\cite{NA49-Alt08}. The distributions can be fitted with a sum of two Gaussian functions
placed symmetrically around mid-rapidity
\be
f(y,a,\sigma)=\frac{1}{\langle N\rangle}\frac{{\rm d} N}{{\rm d} y}
=\frac{1}{\sqrt{8\,\pi\, \sigma^2}}\, \Big[
e^{-\frac{(y-a)^2}{2\, \sigma^2}}+e^{-\frac{(y+a)^2}{2\, \sigma^2}}
\Big]\,,
\label{2Gauss}
\ee
\begin{figure}
\centerline{
\parbox{7cm}{\includegraphics[width=7cm]{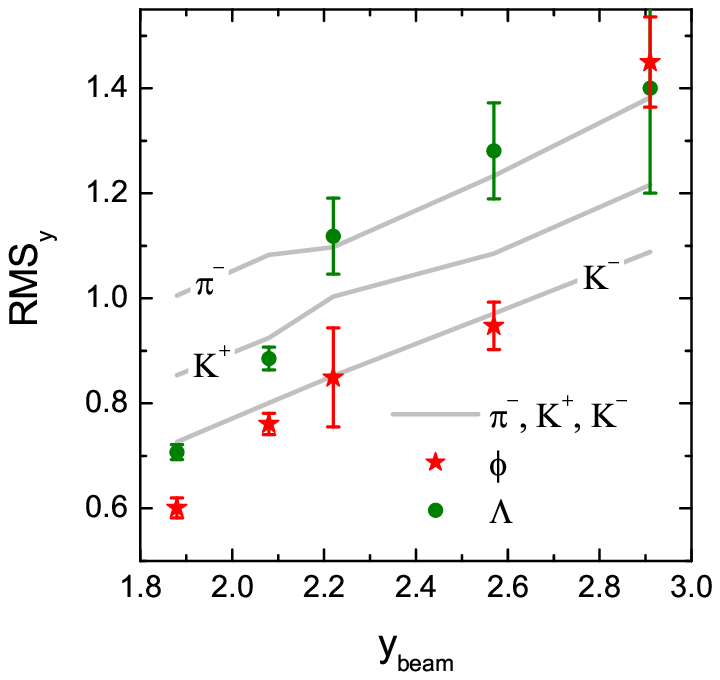}}
\quad
\parbox{7cm}{\includegraphics[width=7cm]{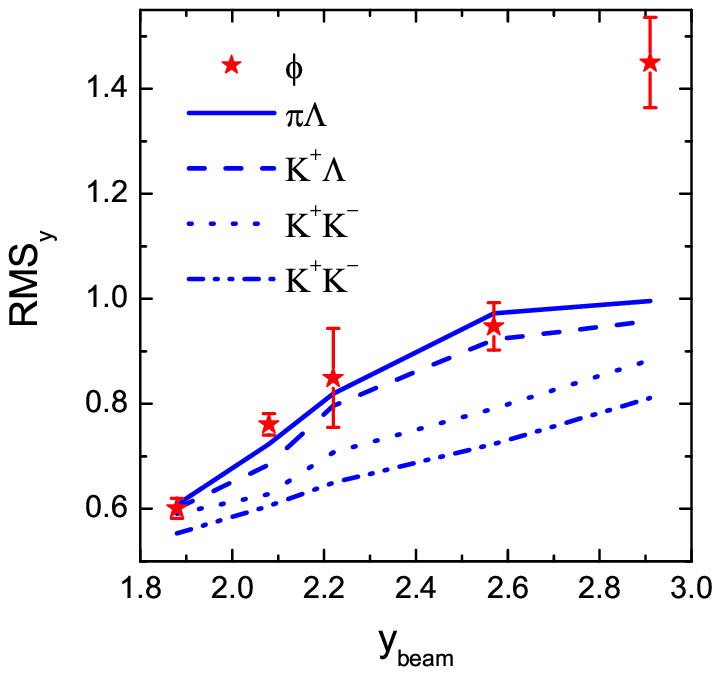}}
}
\caption{Root mean square of the rapidity distributions of particles produced in Pb+Pb collisions
at SPS energies versus the beam rapidity. Left panel:
experimental data for $\pi^-$ and $K^\pm$ from ref.~\cite{NA49-Afanasiev02,NA49-Alt08-piK},
$\phi$ mesons from~\cite{NA49-Alt08} and $\Lambda$ from~\cite{NA49-Alt08-Lb}. Right panel:
The distribution width for $\phi$ mesons produced in $\pi\Lambda\to \phi Y$, $K^+\Lambda\to \phi N$ and
$K^+K^-\to \phi$ reactions calculated with eq.~(\ref{rms12}), except the dash-dot-dot line
calculated with eq.~(\ref{Gauss-RMS}) in comparison with the experimental data from~\cite{NA49-Alt08}. }
\label{fig:Rap}
\end{figure}
here $y$ is the rapidity in c.m system. The width of the distribution is characterized by the root
mean square $\mbox{\sf RMS}^2=\sigma^2+a^2$\,. The left panel of figure~\ref{fig:Rap} depicts the
$\mbox{\sf RMS}$s versus the beam rapidity for $\phi$ mesons, constructed from the parameters
$\sigma$ and $a$ given in table~V in~\cite{NA49-Alt08}, together with those for $\pi^-$, $K^+$ and
$K^-$~\cite{NA49-Afanasiev02,NA49-Alt08-piK}. Ref.~\cite{NA49-Alt08} pointed out that the width
of the $\phi$ meson distribution does not fit into the systematics observed for the other mesons but
increases much faster with the collision energy. It was emphasized in~\cite{NA49-Alt08} that the steep rise of
the $\phi$ distribution width cannot be explained by the hadronic process $K^+\, K^-\to \phi$. The
rapidity distribution of $\phi$ from the latter process can be estimated as the product of $K^+$
and $K^-$ rapidity distributions. The width of the product was calculated in~\cite{NA49-Alt08}
under the assumption of the single-Gaussian sources for kaons with $\mbox{\sf RMS}_{K^+}$ and
$\mbox{\sf RMS}_{K^-}$ as
\be
\mbox{\sf RMS}_{K^+K^-}=\frac{\mbox{\sf RMS}_{K^+}\, \mbox{\sf RMS}_{K^-}}
{\sqrt{\mbox{\sf RMS}^2_{K^+}+\mbox{\sf RMS}^2_{K^-}}}\,.
\label{Gauss-RMS}
\ee
We show this result on the right panel of figure~\ref{fig:Rap} by the dash-dot-dot line. The resulting
width is much smaller than that observed in the experiment.
Question is if any other $\phi$ production mechanism can produce the steep rise of the
$\mbox{\sf RMS}$ with the collision energy. We note that the rapidity distribution of hyperons
increases much faster than those for mesons as the hyperons are dragged with nucleons to forward
and backward rapidities (an effect of partial transparency). The $\mbox{\sf RMS}$s of $\Lambda$
reported in ref.~\cite{NA49-Alt08-Lb} are shown on the left panel in figure~\ref{fig:Rap} by solid circles. The distribution width for
158~$A$GeV was not determined experimentally, therefore we calculated it from the rapidity
distribution obtained theoretically in~\cite{Ivanov06} within a hydrodanamic model which
successfully reproduces the particle production.

We assume that the rapidity distributions of particles do not change after some initial stage when
nuclei are passing throuhgh each other. This implies the absence or weakness of acceleration and
diffusion processes. The collision kinematics is restricted mainly to the exchange of
transverse momenta. Then the rapidity distribution of $\phi$s produced in
the reaction $1+2\to \phi +X$  is roughly
proportional to the product of rapidity distributions of colliding particle species 1 and 2.
If the distributions are given by the two Gaussian distributions (\ref{2Gauss}) the {\rm RMS}
of the resulting $\phi$ distribution is given by
\be
\mbox{\sf RMS}^2_{12}=\frac{\dsp\intop_{-\infty}^\infty d y\, y^2\,f(y,a_1,\sigma_1)\,f(y,a_2,\sigma_2)}
{\dsp\intop_{-\infty}^\infty d y\,f(y,a_1,\sigma_1)\,f(y,a_2,\sigma_2)}
\nonumber\\
= \frac{\sigma_1^2\,\sigma_2^2}{\sigma_1^2+\sigma_2^2}+
\frac{a_1^2\sigma_2^4+a_2^2\,\sigma_1^4}{(\sigma_1^2+\sigma_2^2)^2}+
\frac{2\,a_1\, a_2\,\sigma_2^2\,\sigma_1^2}{(\sigma_1^2+\sigma_2^2)^2}\,
\tanh\frac{a_1\,a_2}{\sigma_1^2+\sigma_2^2}\,.
\label{rms12}
\ee
This expression differs from (\ref{Gauss-RMS}) applied in~\cite{NA49-Alt08}. Using the parameters
of $K^+$ and $K^-$ from~\cite{NA49-Afanasiev02,NA49-Alt08-piK} and eq.~(\ref{rms12}) we obtain
{\sf RMS}s for the  $K^+\, K^-\to\phi$ reaction shown in figure~\ref{fig:Rap} (right panel) by
dotted line. They are somewhat larger than those obtained in~\cite{NA49-Alt08} but still much
smaller than empirical ones. In contrast, the width of $\phi$ rapidity distributions from
the reactions involving $\Lambda$ particles,  $\pi\Lambda\to \phi Y$ and $K^+\Lambda\to \phi N$
(solid and dashed lines in figure~\ref{fig:Rap}) rises much faster and is comparable with the
experimental results for beam energies between 20~$A$GeV and 80~$A$GeV. In collisions at
158~$A$GeV the $\phi$ distribution is much broader than our estimates. Perhaps, some new
non-hadronic mechanism of the $\phi$ production becomes operative at this energy.

\section{Conclusions}

In this paper we study a new mechanism of $\phi$ meson production in nucleus-nucleus
collisions---the catalytic reactions on strange particles, e.g., $\pi Y\to \phi Y$ and $\bar{K}
N\to \phi N$. These reactions are OZI-allowed and their cross section can be by an order of
magnitude larger than the cross section of conventional OZI-suppressed $\phi$ production reactions
$\pi N\to \phi N$ and $N\Delta\to\phi N $, etc, considered so far. These reactions require only
one strange particle in the entrance channel and therefore are less suppressed than the
strangeness coalescence reactions, $K\bar K\to \phi$ and $KY\to \phi N$ in collisions where
strangeness is statistically suppressed. Using a hadronic Lagrangian~(\ref{lag}) we estimate the
$\pi Y\to \phi Y$ and $\bar{K} N\to \phi N$ cross sections to be roughly of the order of 1~mb. In
order to estimate the efficiency of the new reactions we calculate the evolution of the
strangeness content of the fireball within a hadrochemical model. The main assumption we follow is
that  kaons ($K^+$ and $K^0$) can leave the fireball freely after being created whereas the
strange particles (hyperons, $K^-$ and $\bar{K}^0$) remain in the fireball in thermal and
chemical equilibrium. Thus the net strangeness accumulated in the fireball is growing in the
course of the collision. We compare the rates of $\phi$ production in the catalytic reactions and
in the $\pi N\to \phi N$ reaction. The former can be competitive in collisions with the maximal
temperature above 110~MeV and the collision time $\gsim 10$~fm. The efficiency of catalytic
reaction increases if some strangeness is presented in the fireball right in the beginning after
the first most violent nucleon-nucleon collisions and if the fireball lifetime is longer.

We discuss how the catalytic reaction could affect the centrality dependence of the $\phi$ yield.
Since the catalytic rates depend on the concentration of strange particles accumulated during some
time, the resulting number of $\phi$s produced in such a reaction would grow $\propto t_0^2$, where $t_0$
is the collision time, in contrast to the number of $\phi$s from conventional reactions $\pi N\to
\phi N$ growing $\propto t_0$\,. In the scaling regime of hydrodynamics the collision time can be
related to the typical length scale of the system  determined by its volume $t_0\propto
l\propto V^{1/3}\propto N_{pp}^{1/3}$, where  $N_{pp}$ is the centrality criterion -- the mean
number of projectile participants. Taking two types of processes into account we can parameterize the
experimental centrality dependence of the $N_\phi/N_\pi$ and $N_\phi/N_{K^+}$ ratios
measured in Au+Au collision at 11.7~$A$GeV as follows
\be
\frac{N_\phi}{N_\pi}\times 10^3 &= & 7\, \Big(\frac{N_{\rm pp}}{A}\Big)^{1/3}
+ 3\, \Big(\frac{N_{\rm pp}}{A}\Big)^{2/3}\,,
\nonumber \\
\frac{N_\phi}{N_{K^+}}\times 10^2 &=& 1.7 + 1.4\, \Big(\frac{N_{\rm pp}}{A}\Big)^{1/3}\,,
\qquad A=197\,.
\nonumber
\ee
Here the first terms are due to the conventional reactions involving non-strange particles and the
second terms are due to the catalytic $\phi$ production on strange particles.
Relative strength of two terms shows that the contribution of the catalytic reactions can be up to
30\%--40\% at the AGS energies for $N_{\rm pp}=A$.

Analyzing the $\phi$ rapidity distributions at SPS energies we find that the strong rise of the
distribution width with the collision energies between 20~$A$GeV and 80~$A$GeV can be explained by
the $\phi$ production in $\pi\Lambda$ and  $K^+\Lambda$ collisions.

The present analysis of the catalytic mechanisms of the $\phi$ production is only exploratory and
should merely serve as an invitation for further investigations. Particularly, a more reliable
calculation of the cross sections for the $\pi Y\to \phi N$ and $\bar{K} N\to \phi N$ reactions
and fireball evolution is mandatory.

\ack
We gratefully acknowledge the support by VEGA under Nr. 1/4012/07.


\section*{References}

\begin{thebibliography}{99}

\bibitem{AGS-Akiba96} Akiba Y {\it et al} (E802 Collaboration) 1996 {\it Phys. Rev. Lett.} {\bf 76} 2021-4
\bibitem{AGS-Back04} Back B B {\it et al} (E917 Collaboration) 2004 {\it Phys. Rev.} C {\bf 69} 054901 (10pp)
\bibitem{NA49-Afanasiev00} Afanasiev S V {\it et al} (NA49 Collaboration)  2000 {\it Phys. Lett. } B {\bf 491} 59-66
\bibitem{NA50-Alessandro03} Alessandro B {\it et al} (NA50 Collaboration) 2003 {\it Phys. Lett. } B {\bf 555} 147-55
\bibitem{CERES-Adamova06} Adamova D {\it et al} (CERES Collaboration) 2006 {\it Phys. Rev. Lett.}
{\bf 96} 152301 (4pp)
\bibitem{NA60-DeFalco06} De Falco A {\it et al} (NA60 Collaboration) 2006 {\it Nucl. Phys.} A {\bf 774}
719c-22c
\bibitem{NA49-Alt08} Alt C {\it et al} (NA49 Collaboration) 2008 {\it Phys. Rev.} C {\bf 78} 044907 (15pp)
\bibitem{STAR-Adams05} Adams J {\it et al} (STAR Collaboration) 2005 {\it Phys. Lett. } B {\bf 612} 181-9
\bibitem{PHENIX-Adler05} Adler S S {\it et al} (PHENIX Collaboration) 2005 {\it Phys. Rev.} C {\bf 72} 014903 (23pp)
\bibitem{FOPI-Mangiarotti03} Mangiarotti A {\it et al} (FOPI Collaboration) 2003 {\it Nucl. Phys.} A {\bf 714} 89-123

\bibitem{NA49-Friese97} Friese V for NA49 Collaboraion 1997 {\it J. Phys. G: Nucl. Part. Phys.} {\bf 23} 1837 (4pp)
\bibitem{NA50-Jouan08}  Jouan D for NA50 Collaboraion 2008 {\it J. Phys. G: Nucl. Part. Phys.} {\bf 35} 104163 (4pp)
\bibitem{NA60-Floris08} Floris M for NA60 Collaboraion 2008 {\it J. Phys. G: Nucl. Part. Phys.} {\bf 35} 104054 (4pp)

\bibitem{Johnson01} Johnson S C, Jacak B V and Drees A 2001 E{\it ur. Phys. J.} C {\bf 18} 645-9
\bibitem{Filip01} Filip P and Kolomeitsev E E 2001 {\it Phys. Rev.} C {\bf 64} 054905 (8pp)
\bibitem{Kolom02} Kolomeitsev E E and Filip P 2002 {\it J. Phys. G: Nucl. Part. Phys.} {\bf 28} 1697-1705
\bibitem{Bleicher03} Bleicher M 2003 {\it Nucl. Phys.} A {\bf 715} 85c-94c
\bibitem{Sibirtsev06} Sibirtsev A, Haidenbauer J and Mei\ss{}ner U-G 2006 {\it Eur. Phys.} A {\bf 27} 263-8

\bibitem{Lipkin76} Lipkin H 1976 {\it Phys. Lett. } B {\bf 60} 371-4

\bibitem{Chung97} Chung W S, Li G Q and Ko C M 1997 {\it Phys. Lett.} B {\bf 401} 1-8
\bibitem{Titov00} Titov A, K\"ampfer B and Reznik B L 2000 {\it Eur. Phys. J.} A {\bf 7} 543-57

\bibitem{Titov02} Titov A, K\"ampfer B and Reznik B L 2002 {\it Phys. Rev.} C {\bf 65} 065202 (14pp)

\bibitem{Doring08} D\"oring M, Oset E and Zou B S 2008 {\it Phys. Rev. } C {\bf 78} 025207 (10pp)

\bibitem{Shor85} Shor A 1985 {\it Phys. Rev. Lett} {\bf 54} 1122-5

\bibitem{Ko91} Ko C M and  Sa B H {\it Phys. Lett.} B {\bf 258} 6-10

\bibitem{Bleicher99} Bleicher M{\it et al} 1999 {\it J. Phys. G: Nucl. Part. Phys.} {\bf 25} 1859-96

\bibitem{Herrmann96} Herrmann N for FOPI Collaboration 1996 {\it Nucl. Phys.} A {\bf 610} 49c-62c

\bibitem{Chung97-2} Chung W S , Li G Q and Ko C M 1997 {\it Nucl. Phys.} A {\bf 625} 347-71

\bibitem{Kampfer02} K\"a{}mpfer B, Kotte R, Hartnack C and  Aichelin J 2002
{\it J. Phys. G: Nucl. Part. Phys.} {\bf 28} 2035-40

\bibitem{Barz02} Barz H W, Z\'et\'enyi M, Wolf Gy  and K\"ampfer B 2002 {\it Nucl. Phys.} A {\bf
705} 223-35

\bibitem{Herrmann08} Herrmann N for FOPI Collaboration 2008 {\it talk at Int. Conf. Strangeness in Quark
Matter 2008, October 6-10, Bejing China}

\bibitem{Agakichev09} Agakishiev G {\it et al}  (The HADES Collaboration) 2009 arXiv:0902.3487 [nucl-ex]


\bibitem{Andronic06}
Andronic A, Braun-Munzinger P and Stachel J 2006 {\it Nucl. Phys.} A {\bf 772} 167-99

\bibitem{lk02}
Lutz M F M and Kolomeitsev E E 2002 {\it Nucl. Phys.} A {\bf 700} 193-308

\bibitem{LL}
Berestetskii V B, Lifshitz E M, Pitaevsky  L P 1982
{\it Quantum Electrodynamics}  (Oxford: Clarendon Pres)

\bibitem{Sibirtsev97}
Sibirtsev A, Cassing W, and  Mosel U 1997 {\it Z. Phys.} A {\bf 358} 357-67

\bibitem{Dahl67}
Dahl O I, Hardy L M, Hess R I, Kirz J and Miller D H 1967 Phys. Rev. {\bf 163} 1377-429

\bibitem{Courant77}
Courant H, Makdisi Y I, Marshak M L, Peterson E A, Ruddick K and Smith-Kintnerl J
1977 {\it Phys. Rev.} D {\bf 16} 1-8

\bibitem{Ko88}
Ko C M and Xia L H 1988 {\it Phys.\ Rev.} C {\bf 38}  179-83

\bibitem{Barz88}
Barz H-W, Friman B L, Knoll J and Schulz H 1988, {\it Nucl.~Phys.} A {\bf 484} 661-84

\bibitem{Barz90}
Barz H-W, Friman B L, Knoll J and Schulz H 1990 {\it Nucl.~Phys.} A {\bf 519} 831-46

\bibitem{Russkikh91}
Russkikh V~N {\it Sov.\ J.\ Nucl.\ Phys.} 1991 {\bf 53} 1037

\bibitem{Russkikh92}
Russkikh V~N and Ivanov Y B 1992 {\it Nucl.\ Phys.}  A {\bf 543} 751-66

\bibitem{Ko83}
Ko C M 1983 {\it Phys. Lett.} B {\bf 120} 294; Ko C M 1984 {\it Phys. Lett.} B {\bf 138} 361

\bibitem{Kolomeitsev95}
Kolomeitsev E E, Voskresensky D N and  K\"ampfer B 1995 {\it Int. J. Mod. Phys.} E {\bf 5} 316-28

\bibitem{Tomasik05}
Tomasik B and Kolomeitsev E E 2005 nucl-th/0512088;\\
Tomasik B and Kolomeitsev E E 2007 {\it Eur. Phys. J.} C {\bf 49} 115-20

\bibitem{Bondorf78}
Bondorf J P, Garpman S I and Zimanyi J 1978 {\it Nucl.\ Phys.} A {\bf 296} 320

\bibitem{Cugnon84}
Cugnon J and Lombard R M 1984 {\it  Nucl. Phys.} A {\bf 422} 635-53

\bibitem{Tsushima94}
Tsushima K, Huang S W and Faessler A 1994 {\it Phys.\ Lett.}  B {\bf 337} 245-53

\bibitem{Tsushima97}
Tsushima K, Huang S W and Faessler A 1997 {\it Austral.\ J.\ Phys.} {\bf 50} 35 (Reprint nucl-th/9602005)

\bibitem{Cassing97}
Cassing W, Bratkovskaya E L, Mosel U, Teis S and Sibirtsev A 1997 {\it Nucl. Phys.} A {\bf 614}
415-32

\bibitem{Tsushima99}
Tsushima K, Sibirtsev A, Thomas A W and Li G Q 1999 {\it Phys. Rev.} C {\bf 59} 369-87

\bibitem{NA49-Afanasiev02} Afanasiev S V {\it et al} (NA49 Collaboration) 2002 {\it Phys. Rev.} C {\bf
66} 054902 (9pp)

\bibitem{NA49-Alt08-piK} Alt C {\it et al} (NA49 Collaboration) 2008 {\it Phys. Rev.} C {\bf
77} 024903 (9pp)

\bibitem{Russkikh92} Russkikh V N and Ivanov Yu B 1992 {\it Nucl. Phys.} A {\bf 543} 751-66

\bibitem{NA49-Alt08-Lb} Alt C {\it et al} (NA49 Collaboration) 2008 {\it Phys. Rev.} C {\bf 78}
034918 (15pp)

\bibitem{Ivanov06} Ivanov Yu B, Russkikh V N and Toneev V D 2006 {\it Phys. Rev.} C {\bf 73}
044904 (29pp)
\end{thebibliography}
\end{document}